\documentclass{appolb}
\usepackage{epsf}

\def\Tr{\mbox{\rm Tr}}

\newcommand{\nnb}{\nonumber}
\newcommand{\bea}{\begin{eqnarray}}
\newcommand{\eea}{\end{eqnarray}}

\newcommand{\barl}{\begin{array}{rl}}
\newcommand{\barr}{\begin{array}{rr}}
\newcommand{\ball}{\begin{array}{ll}}
\newcommand{\ea}{\end{array}}
\newcommand{\APA}{{\tt APACIC++} }
\newcommand{\AME}{{\tt AMEGIC++} }
\newcommand{\PYT}{{\tt PYTHIA} }
\newcommand{\HER}{{\tt HERWIG} }
\begin{document}

\eqsec  
\title{APACIC++, A PArton Cascade In C++}
\thanks{Invited talk presented at the 39. Cracow Summer School
        of Theoretical Physics}

\author{F.~Krauss
\address{Department of Physics, Technion, Haifa 32000, Israel}
\and	
R.~Kuhn
\address{Institut f\"ur Theoretische Physik, TU Dresden 
         01062 Dresden, Germany\\
         Max--Planck--Institut f{\"u}r Physik Komplexer Systeme,
         01187 Dresden, Germany}
\and
G.~Soff
\address{Institut f\"ur Theoretische Physik, TU Dresden 
         01062 Dresden, Germany}
}
\maketitle

\begin{abstract}
In this talk the newly developped Monte--Carlo event generator
{\tt APACIC++} suitable to describe multijet--events in high--energetic
electron--positron annihilations is presented. A new ansatz to
match the corresponding matrix elements for the production of jets
via the strong and electroweak interactions to the subsequent
parton shower modelling the inner--jet evolution is discussed in some
detail. Results obtained with {\tt APACIC++} are compared to other
QCD event generators and to some representative experimental data.
\end{abstract}

\begin{center}
\PACS{13.65.+i, 13.87.-a}
\end{center}
%
%
\section{Introduction}

For decades, electron--positron collisions have been an extensively used
testing ground for quantum field theory and particle physics. Especially
$e^+e^-$--annihilations into hadrons at high energies proved to be of 
continuos interest. In principle such processes can be reliably described 
with the help of Monte--Carlo approaches in the form of so--called event 
generators. There, the description of $e^+e^-$--annihilations into hadrons 
can be divided into three steps. First, a number of
partons is produced at a scale of the order of the c.m. energy of the incoming
electron--positron pair. Here, the standard method of perturbative quantum
field theory of summing and squaring amplitudes related to corresponding
Feynman--diagrams is applicable. In a second step, these primary partons
loose virtual mass and energy by radiating additional partons giving rise 
to jets. Because of the possibly high and varying number of particles 
involved here one has to abandon the idea of summing the full amplitudes. 
Instead one considers only the limits of soft and small angle emissions 
resulting in a probabilistic description of jet--evolution as a chain of 
nearly independent single emissions in the perturbative regime of strong 
interactions. These radiations stop in a third step at some infrared scale 
of the order of a few $\Lambda_{\rm QCD}$ and hadronization sets in. Since 
this is essentially a non--perturbative, soft process it is usually modelled
by some parameter--dependent phenomenological hadronization scheme, which 
does not alter the density-- and energy--distribution of particles in 
phase--space drastically. However, the parameters entering the model are to 
a large extent scale--dependent. Therefore the jet--evolution via the parton 
shower has the additional purpose to connect the high--energy scale of
jet--production with the low--energy scale of hadronization and thus 
guarantees the universality of the hadronization scheme once the parameter
are fixed to fit the data. 

Monte--Carlo event generators are perfectly capable to model high--\-energetic
$e^+e^-$--annihilations into hadrons by means of the three steps as
described above. In this respect, they are an indispensable tool to bridge
the gap between theoretical considerations concering the dynamics of such 
events and their experimental observation and to provide testable
signatures in a well--defined manner.

With rising energies, however, an increasing number of particles and of
jets is produced, and the production, observation and theoretical description
of such multijet--events is one of the cornerstones of current particle 
physics. Various reasons feed this interest and we would like to highlight
only briefly some of them.

First of all, large parts of what is known as the Standard Model has been 
tested via multijet--events. QCD has been established as the correct gauge
theory underlying the strong interactions by means of measurements 
\cite{4JETth} of the Casimir--operators \cite{FIE89}

\bea
C_F = \frac{N_c^2-1}{2 N_c} = \frac{4}{3}\;,\;\;
C_A = N_c = 3
\eea

of the fundamental and adjoint representation and by measuring the
overall normalization

\bea\label{Norm1}
\Tr\left[T^aT^b\right] = \delta^{ab}\,T_F = \frac12
\eea

of its generators. In addition, the electroweak sector of the Standard
Model has been tested exhaustively by precision measurements of observables
like for instance the widths of the gauge bosons and by establishing the 
non--abelian structure of the gauge group via proving the existence 
of the triple gauge boson vertices.

Second, multijet--events open the door to new physics. For example,
the Higgs--boson of the Standard Model has some clear signatures in
$e^+e^-$--collisions in so--called Higgsstrahlungs--processes \cite{Hstrahl}
resulting
in at least four final state fermions which may form jets. Of course,
there is a large variety of other interesting signatures connected 
with cross--sections which depend sensitively on the number $n_f$ of
active flavours in the case of strong interacting particles \cite{GLUINO}.

Therefore it is of some interest, to have at hand some event generator
capable to deal with such multijet--events. One of the major obstacles on that
road is the question of how to match the parton shower responsible for
jet--evolution and going down to the hadronization--scale with the
matrix elements describing the high--energetic production of the jets, since
only this guarantees the universality of the hadronization--scheme used and 
hence the predicitive power of the event generator. Within \APA we have 
implemented a new ansatz to that question enabling this code to deal with
multijet--events due to the strong or electroweak interactions. So the outline
of this article is as follows: In Section II we would like to introduce
briefly some concepts and tools related to the perturbative treatment
of jet--production via matrix elements and their implementation in \APA. 
The parton shower picture of jet--evolution and the way \APA handles it is 
discussed in Section III. There, we briefly compare matrix elements and the
parton shower and their regimes of validity. In Section IV we will comment 
on two main approaches to the question of matching. Additionally, we will 
discuss in some length the ansatz used by \APA. We want to justify this 
ansatz in Section V by considering some results of \APA and comparing them 
to experimental data and the results obtained from other event generators. 
Finally, we would like to conclude in Section VI.

\section{Jet--production in \APA}

\subsection{General features}

Usually, the hadrons produced in $e^+e^-$--annihilations are clustered in 
jets, objects separated in phase--space by some jet--measure. Popular
jet--measures are the JADE-- \cite{JAD} and the DURHAM--scheme \cite{DUR},
defined by

\bea\label{jetschemes}
\barl
(p_i+p_j)^2 = 2E_iE_j\,(1 - \cos\theta_{ij}) > y_{\rm cut}\, s^{(0)}_{ee}
             &\mbox{\rm (JADE)}\\[2mm]
2\,{\rm min}\{ E_i^2,E_j^2\}\,(1 - \cos\theta_{ij}) > y_{\rm cut}\, 
s^{(0)}_{ee} &\mbox{\rm (DURHAM)}
\ea
\eea

for two massless particles to belong to different jets. The parameter 
$y_{\rm cut}$ is a measure for the hardness of the jet. Within perturbation 
theory, the emergence of jets is described by the appropriate matrix elements 
for $e^+e^-\to n\,\mbox{\rm partons}$ thus identifying jets with hard produced 
partons. Evaluating the corresponding cross sections in the standard way by 
squaring amplitudes and integrating over the phase space available one is,
even at the tree--level, confronted with divergencies. Beyond the tree--level 
more divergencies occur due to additional loops or legs and have to be
treated. Here, we would merely like to state, that mutual cancellations 
of the divergencies due to loops and legs connect topologies of varying numbers
of legs and pose a major obstacle to any calculation beyond the tree--level.
Some of the recent results can be found for instance in \cite{6JETLO,4JETNLO}.

However, at the moment \APA deals with tree--level matrix--elements only.
They can be kept finite quite easily by merely subjecting the initial partons 
to the condition that they form well--separated jets, i.e. by applying the
restrictions of Eq.\ref{jetschemes} to the integration over the final--state
phase--space. It is not much of a surprise that the corresponding jet cross 
sections now are finite in Leading Order and become divergent for 
$y_{\rm cut} \to 0$. 

Consequently a choice of this initial $y_{\rm cut}$ yields a parametrization 
of the reliability of LO matrix elements, and softer parton emissions are 
supposed to be better described by the appropriate Sudakov form factor, see
Section III. To summarize, this treatment is nothing else but the statement, 
that \APA considers jets to be entities whose production can be described in a 
reliable and controllable manner by traditional perturbation theory, i.e. by 
matrix elements.

Some of the efffects of higher order QCD--corrections are implemented in
\APA by an overall factor $\kappa_s<1$ for the scale of the strong coupling 
constant. Some similar treatment can be found for instance in
\cite{PYTHIA,SPINORS2}. This factor is a fit parameter for the scale of 
$\alpha_s$ used within the matrix elements in the form

\bea
\alpha_s^{\rm M.E.} = \alpha_s(\kappa_s s)\,.
\eea

\APA uses common the LO running of $\alpha_s$ and the quark masses (see for 
instance \cite{FIE89,ESW96}), where the scale of the latter ones is not 
affected by $\kappa_s$.

\subsection{Defining relative rates}

Within \APA there are matrix elements for the production of two and three
QCD--jets via the exchange of a photon or a $Z$. Denoting the 
cross--sections by $\sigma_{q\bar q}$ and $\sigma_{q\bar qg}$, respectively,
the corresponding rates are given by 

\bea
\label{two-three}
{\cal R}_3 = \sigma_{q\bar qg}/\sigma_{q\bar q}\;,\;\;\; 
{\cal R}_2 = 1-{\cal R}_3\,,
\eea

based on the probabilistic picture of 3--jets being an exclusive subset
of the inclusive production of hadrons \cite{ESW96}. When dealing with higher
numbers of jets produced by QCD only, one is left with the task to extend
this scheme in a sensible manner. Within \APA we provide at least three 
schemes, namely a ``direct'' scheme, and two ``rescaled'' ones, which we
denote by ``rescaled1'' and ``rescaled2''

\bea\label{Rates}
\nonumber
{\cal R}_n^{\rm dir.} &=& 
\frac{\sigma^{\rm tot}_n}{\sigma_{q\bar q}}\,,\\
\nonumber
{\cal R}_n^{\rm res1.} &=& \frac{\sigma^{\rm tot}_n}{\sigma_{q\bar q}}\cdot
\prod\limits_{m>n}\left(1-{\cal R}_m^{\rm re.}\right)\quad\quad\mbox{\rm or}\\
{\cal R}_n^{\rm res2.} &=& {\cal R}_n^{\rm dir.}-{\cal R}_{n+1}^{\rm dir.}\,,
\eea

where the last one uses the direct rates ${\cal R}_n^{\rm dir.}$ and the
corresponding rescaling applies for $n<n_{\rm max}$. The related rate 
${\cal R}_{n_{\rm max}}$ remains unchanged. 

It has to be stressed here, that these schemes are obviously by no means
consistent in $\alpha_s$, i.e. perturbation theory. Instead the evaluation of
the rates and consequently the admixture of different jet--numbers within
\APA is to some extent just a phenomenological model with $\kappa_s$ the
parameter to be fitted to data. 

Of course the situation above with QCD only changes drastically taking into 
account the production of jets via more than one electroweak gauge boson, for
example when considering four--jet production via $W$--, $Z$-- or 
Higgs--bosons beyond the corresponding thresholds. Currently this situation is 
handled in the following way. If the electroweak production of four or 
more jets is taken into account, the cross sections and the corresponding 
rates are divided into two sets. The first subset is defined by four or more 
fermions in the final state (electroweak subset), the second set is the 
conjugate subset (QCD subset). Interferences occuring between both of them,
for example if an internal $Z$-- or $\gamma$--line is replaced by a gluon,
are awarded to the electroweak set. Then the rate of the first set is obtained 
by the sum of the corresponding cross sections and the rate of the second set 
still is defined via the cross section for the inclusive production of a 
quark--antiquark pair. Within the electroweak subset the single rates are 
determined by the appropriate cross sections, within the QCD subset the 
determination of the relative rates is achieved in the fashion of 
Eq.\,(\ref{Rates}).

\subsection{Multijet--matrix elements available}

To allow for the formation of higher jet--configurations we have added
three matrix element generators.

\begin{enumerate} 
\item{In its present state, \AME \cite{AMEGIC} describes the production of up 
to five massive jets via the strong or electroweak interaction in Leading 
Order. 

Recently, the production of up to five jets via the strong interaction has 
been successfully tested. Results obtained by \cite{SPINORS2} in this channel 
have been reproduced for both massless and massive quarks and over the full 
ranges of the two jet--measures considered, namely the JADE-- and the 
DURHAM--scheme. Additionally, the production of four jets
by the exchange of two electroweak gauge bosons ($W$, $Z$ or $\gamma$) has been
tested by reproducing some of the results of \cite{EXCALIBUR}.}
\item{{\tt DEBRECEN} \cite{DEBRECEN} accounts for the QCD--production of up
to 5 jets in Leading Order and up to 4 jets in Next-to Leading Order.}
\item{{\tt EXCALIBUR} \cite{EXCALIBUR}
describes processes with 4 quark--jets, generated 
via strong or electroweak interactions in Leading Order.}
\end{enumerate}

One of them, \AME, has not been published yet. In its final version it is 
meant to allow for the production of up to six massive jets in all Standard 
Model channels including the full electroweak and Higgs--sectors. 
\AME uses the helicity method of \cite{SPINORS2} originally proposed in
\cite{SPINORS1}. 

For all of the matrix element generators \APA provides interfaces.
In addition \APA and \AME, respectively, allow for the inclusion of 
QED--Coulomb corrections to the production of heavy particles near the
threshold \cite{WWCoul}
and for some initial state radiation of photons off the electrons \cite{ISR1}.

\section{\label{P.S.} Final state parton shower}

\subsection{Space--time picture and LLA}

We would like to dwell on the jet--evolution of the initial partons produced
by the appropriate matrix elements. Here, the common approach of evaluating 
and squaring amplitudes fails due to the high number of particles involved. 
To deal with this, one restricts oneself to the kinematical enhanced regions 
of small angles and low energies of the emitted particles. This allows for the
probabilistic construction of the jet--evolution in terms of subsequent 
independent branchings of one parton into two. The kinematical enhancement is 
a common feature of all field theories with massless bosons, where the 
regions of soft and collinear emissions give rise to the corresponding 
divergencies. To illustrate this point, we would like to consider the 
amplitude squared of a $(N+1)$--particle matrix element 
obtained via one additional radiation from a $N$--particle matrix element 
\cite{ESW96},

\bea\label{Singlebranch}
|{\cal M}_{N+1}|^2 \propto \frac{\alpha_s}{t_a}\,C\,P_{ba}(z)\,
|{\cal M}_N|^2\,,
\eea

where $C$ is some appropriate colour factor, $t_a$ the virtual mass of the 
particle $a$ decaying into $b$ and $c$, and $P_{ba}(z)$ is the corresponding
splitting function depending on the energy fraction $z$ particle $b$ carries 
away. People familiar with the splitting functions will appreciate the fact,
that within the framework of QCD event--generators, the notorious divergencies
related to the limits $z\to 0$ and $z\to 1$ are regularized kinematically
in quite a natural way by imposing some minimal virtual mass for any outgoing
parton. As can be deduced directly from Eq.\,(\ref{Singlebranch}) each decay 
process $a\to bc$ may be described by means of two variables only, namely 
$t_a$ and $z$. Note, that there are different possibilities to interpret 
those two parameters and they refer to different schemes of organizing the 
parton shower to be reviewed later on. To specify the process kinematics of 
the decay $a\to bc$ completely, an additional azimuthal angle $\phi$ of the 
decay plane around the direction $a$ is needed. As a first guess $\phi$ is 
distributed isotropically, but rather weak spin correlations of two subsequent
branching processes lead to some non--trivial plane correlation, which is
included in \APA, too \cite{AZIMUT}.

However, the cross section related to the process of Eq.\,(\ref{Singlebranch})
can now be written as

\bea\label{SingleBWQ}
d\sigma_{N+1} \propto \alpha_s \,dz\frac{dt_a}{t_a}\,\hat P_{ba}(z)\,
d\sigma_N,
\eea

Iterating this equation it is easy to see that a strong ordering of the 
virtual masses related to subsequent emissions yields the largest enhancement
of the form 

\bea
d\sigma_N\propto d\sigma_0\,\alpha_s^n\,
\int\limits_{Q_0^2}^{Q^2}\frac{dt_1}{t_1}\,
\int\limits_{Q_0^2}^{t_1}\frac{dt_2}{t_2}\,\dots\,
\int\limits_{Q_0^2}^{t_{n-1}}\frac{dt_n}{t_n}
= d\sigma_0\,\frac{\alpha_s^n}{n!}\,\left(\log\frac{Q^2}{Q_0^2}\right)^n\,,
\eea

with $Q^2$ the hard scale of the first parton taking part in the
jet evolution and $Q_0^2$ the infrared scale characterizing usually the onset
of hadronization.

This can be plugged into a form suitable for the implementation within a code
by considering first the well--known DGLAP--Equation \cite{DGLAP}

\bea\label{DGLAP1}
t\,\frac{\partial}{\partial t}\,q(x,t)=\int\,dz\,
\frac{\alpha_s}{2\pi}\,P(z)\,\left[\frac{1}{z}\,q\left(\frac{x}{z},t\right)
-q(x,t)\right]
\eea

describing the evolution of a parton density $q(z,t)$ inside a hadron.
Introducing the Sudakov--form factor \cite{SUDAKOV}

\bea\label{Suddef}
\Delta(t,t_0)\equiv\exp{\left\{-\int\limits_{t_0}^t\,\frac{dt'}{t'}
\int\limits_\epsilon^{1-\epsilon}\,dz\,
\frac{\alpha_s}{2\pi}\,P(z)\right\}}\,,
\eea

one is able to construct an evolution equation similar to the DGLAP--equation 
and to rewrite it as an integral equation,

\bea\label{Sudint}
q(x,t)=
\Delta(t,t_0)\,q(x,t_0)\,+\,
\int\limits_{t_0}^{t}\frac{dt'}{t'}\frac{\Delta(t,t_0)}{\Delta(t',t_0)}
\int\limits_\epsilon^{1-\epsilon}\,\frac{dz}{z}
\frac{\alpha_s(p_\perp^2)}{2\pi}\,P(z)\,q\left(\frac{x}{z},t'\right) \,,
\eea

allowing for a probilisitc interpretation of the Sudakov form factor 
$\Delta(t,t_0)$ as the probability, that no branching occurs between 
$t$ and $t_0$. This interpretation is further motivated by the observation,
that 

\bea
\Delta(t_0,t_0) = 1\;\mbox{\rm and}\;\;\;
\Delta(t,t') = \frac{\Delta(t,t_0)}{\Delta(t',t_0)}\,.
\eea

In Eq.\,\ref{Sudint} we have included explicitly the scale of the strong 
coupling constant in terms of the transversal momentum, which is in LLA
given by

\bea
p_\perp^2=z(1-z)t\,.
\eea

As mentioned above, there is some minimal virtual mass for each parton,
$t_0$, regularizing the divergencies via limiting the $z$--integration

\bea\label{zcuts0}
\epsilon(t)\leq z\leq 1-\epsilon(t)\;,\;\;\;
\epsilon(t) = \frac12-\frac12\,\sqrt{1-4\,\frac{t_0}{t}}\,.
\eea  

We have seen, that forcing a strong ordering of the virtual masses results in 
a resummation of the leading logarithms (leading logarithmic approximation, 
LLA). The divergencies related to the singular behaviour of the splitting 
function at the edges of the $z$--space are cut in this approach. Their 
resummation is achieved in the modified leading logarithmic approximation 
(MLLA). Its basic ideas will be discussed below.

\subsection{Coherence effects and MLLA}

\begin{figure}
\begin{center}
\mbox{\epsfxsize=6cm\epsfysize=5cm\epsffile{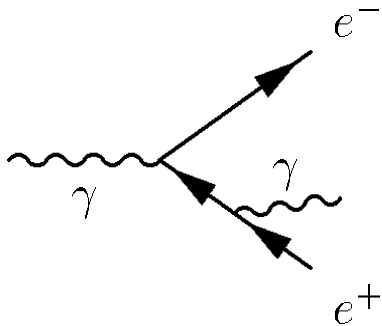}}
\parbox[t]{13.5cm}{\caption{\label{Chud}
         Emission of a photon by an $e^+e^-$--pair in QED.}}
\end{center}
\end{figure}

To illustrate the idea of coherence, we want to refer first to the 
Chudakov--effect of QED \cite{Chu}. Here, an $e^+e^-$--pair is produced 
off an initial virtual photon and emits an additional photon, see 
Fig.\,\ref{Chud}.
Assuming the photon to stem from the positron, the formation time $t_f$ of
this photon can be estimated from the uncertainty principle as

\bea
t_{f}\approx\frac{1}{k\theta_{\gamma e}^2}\approx
\frac{\lambda_\perp}{\theta_{\gamma e}}\;,
\eea

where $k$ is the photon momentum, $\lambda_\perp$ the component of its
wavelength vector transversal to the positron and $\theta_{\gamma e}$
is the positron--photon angle. To allow this photon to resolve the 
electron--positron pair and hence to be produced at all, the distance the 
pair separates during the photon formation should be larger than the 
transversal wavelength. Therefore we end up with

\bea
\rho_\perp\approx
t_f\theta_{ee}\approx\frac{\theta_{ee}}{\theta_{\gamma e}}\,\lambda_\perp
\geq \lambda_\perp\;\;\longrightarrow\;\;
\theta_{\gamma e}\leq\theta_{ee}
\eea

yielding an angular ordering. Stated the other way around, the emission of 
a photon is suppressed at angles larger than $\theta_{ee}$ since it 
experiences only the effect of the overall charge of the pair. A similar 
reasoning applies for QCD and thus motivates the angular ordering of 
subsequent emissions within jets to model coherence effects like the one of 
the example above \cite{MLLA,ESW96}. 

Within \APA, angular ordering of the jet--evolution is accomplished in 
two ways. The first approach is to subject subsequent emissions to a hard veto
on rising angles of subsequent branchings \cite{ANGORD}. The second method 
utilizes the fact, that it is equally possible to construct a Sudakov form 
factor for angular ordering as 

\bea\label{SudMLLA}
\Delta_i(\zeta E^2) = \exp\left[-\int\limits_{4t_0}^{\tilde t}
\frac{dt'}{t'}\int\limits_{\epsilon(\zeta)}^{1-\epsilon(\zeta)}
\,dz\frac{\alpha_s(z^2(1-z)^2t')}{2\pi}\,\hat P_{ji}(z)\right]\,.
\eea

by taking into account not only Leading Logarithmic contributions, but also 
double leading logarithmic terms.  
Note, that neglecting some redefinitions of integration variables and
adjusting the regions of integration this form is exactly the one
of the LLA Sudakov form factor yielding the same interpretation as above.

The new evolution variable is given with the parton's energy $E$ by 

\bea\label{zetadef}
t' = \zeta E^2\;,\;\;\;
\zeta = \frac{p_b\cdot p_c}{E_b E_c} \approx 1-\cos\theta_{bc}\;,\;\;\;  
\epsilon(\zeta) = \sqrt{t_0/(E^2\zeta)}
\eea

and the transversal momentum is now

\bea
p_\perp^2 = z^2(1-z)^2 \,\zeta E^2\,.
\eea

An additional remark is in order here. Since the construction of the MLLA
Sudakov form factor relies on the assumption of small branching angles as can 
be deduced from Eq.\,(\ref{zetadef}), this ordering scheme might not be
applicable to the first branchings within a jet, which can very well
include regions of $\cos\theta_{bc}$ and $\zeta$ yielding a virtual mass 
$\sqrt{\tilde t} = \sqrt{\zeta} E$ of particle $a$ larger than its energy. 
To cure this problem, within \APA the first decay is always performed using 
the LLA form factor. Other codes employ the fact, that $\cos\theta_{bc}$ and 
$\zeta$ are not boost invariant and perform the evolution of the jets in 
suitable reference frames. 

For further details on the construction of the Sudakov form factors in the
two ordering schemes, virtualities (LLA) and angles (MLLA) we refer to
\cite{Dok91} and to the concise textbook \cite{ESW96}.

\subsection{Matrix elements vs. parton shower}

To compare matrix elements and the parton shower and discuss their
regions of reliability, it is sufficient, to stress once again, that the 
construction of the Sudakov form factor and consequently the organization of 
the shower relies on the expansion around the soft and collinear limit 
including proper resummation of the large logarithms attached to each region.
Therefore the parton shower performs better than matrix elements in this 
region. However, vice versa in the region of hard and large--angle emissions 
we should assume the matrix elements to account for a much better description,
since they include interference effects, which become important
when leaving the soft and collinear region.

\section{Matching of matrix elements and the parton shower\label{matchit}}

\subsection{Basic ideas}

We now turn to the question of how to match the matrix elements and the 
parton shower. Actually, this question can be stated in another way, namely 
of how to supply the particles produced in the hard process with virtual
masses to allow them to radiate additional partons. Since the matrix elements
describe the production of on--shell particles only, this question already
suggests an interpretation of the virtual masses as order parameters inside 
the parton shower and as small perturbation not altering anything else.
Before we discuss in some detail the answer to the question above as given 
within \APA we would like to describe briefly the matching algorithms 
employed by other event--generators in the framework of 
$e^+e^-$--annihilations. 

In general, two approaches exist. The first possibility is to utilize the
matrix elements merely to correct the kinematics of the shower evolution of 
the two initial partons with the help of a veto--algorithm
\cite{MATCHING}. This is the 
approach chosen by \PYT \cite{PYTHIA}, where at the present state the first 
radiation according is corrected in a way to reproduce exactly the 
three--jet matrix element. Of course, this accounts for some of the
features of four--jet production as well. Alternatively, one might try to 
divide the phase space for the emission of partons additional to the two 
initial ones in two regions, the hard one dominated by the matrix elements 
and describing the production of further jets, and the soft one modelling 
the inner jet--evolution
\cite{MATCHJET,4MATCH}. The two regions have to be separated, this is to be 
achieved by defining and fitting accordingly a fixed matching scale 
$Q^2$ determining in some sense the virtualities of the out--going partons. 
Consequently, below this scale the parton shower governs the emissions, above 
the matrix elements are responsible. This is the approach chosen within \HER
\cite{HERWIG} in the framework of a MLLA parton shower.

\APA rather follows the second approach of splitting the phase space into two 
regions. But instead of fixing a scale the matching is achieved in a different
way. First of all, we define a $y_{\rm cut}=y_{\rm crit}$ characterizing jets
at the parton level. Then, since for any hard jet production characterized by
$y_{\rm cut}>y_{\rm crit}$ the matrix elements do a better job, they are
responsible for all such emissions. Reversely, the parton shower performs 
better in the soft region characterized by relative low $y_{\rm cut}$.
Therefore the parton shower governs all branchings with 
$y_{\rm cut}<y_{\rm crit}$. Thus within \APA the matching strategy is to use 
the matrix elements for jet--production and the parton shower for their 
evolution. The virtual masses of the outgoing partons are always provided by 
the Sudakov form factor and subjected to the condition that the parton shower 
does not produce any additional jet as specified by $y_{\rm crit}$ \cite{KKS}. 

\subsection{Matching procedure}
\begin{figure}
\begin{center}
\mbox{\epsfxsize=12cm\epsfysize=18cm\epsffile{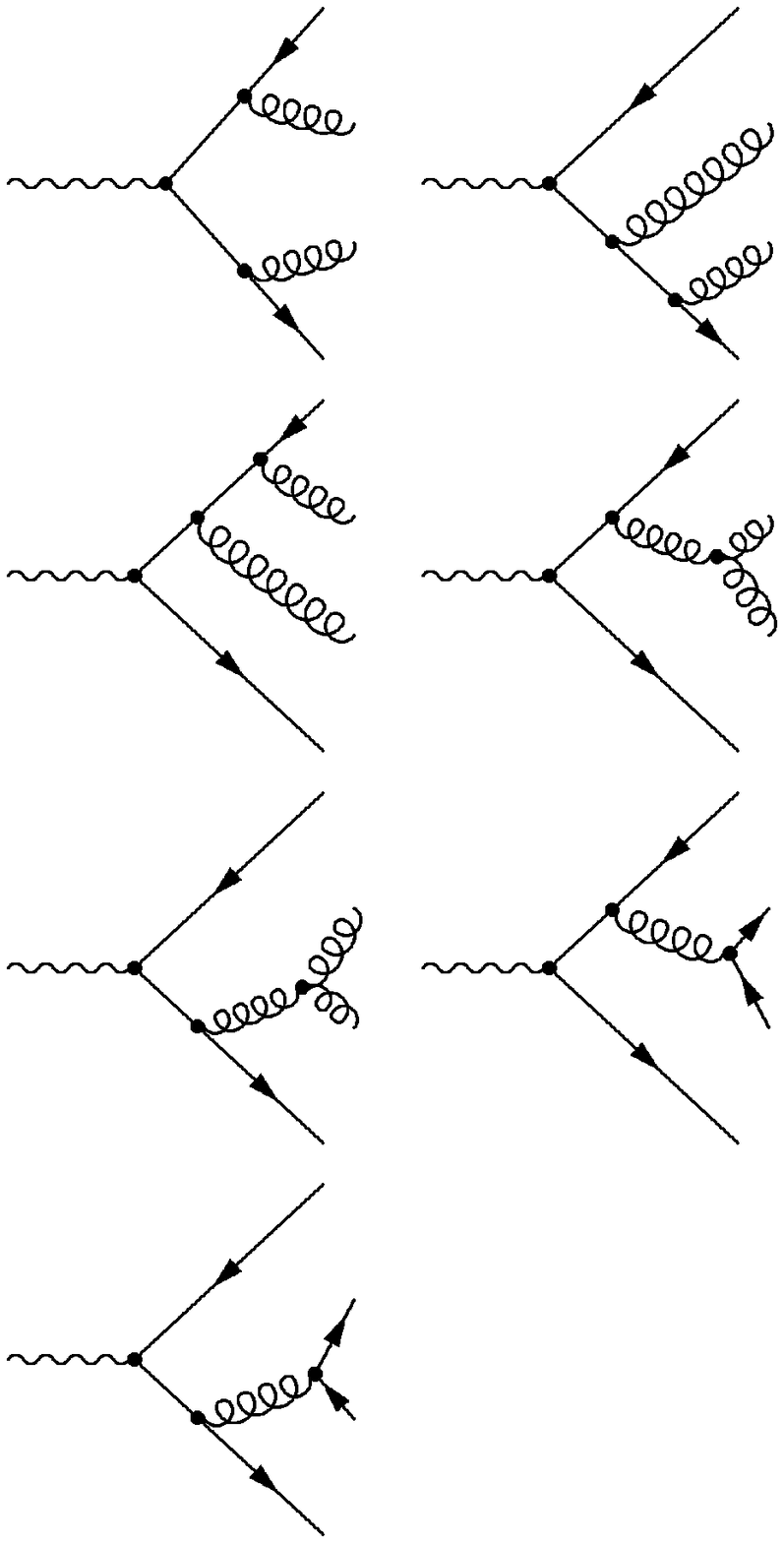}}
\parbox[t]{10cm}{\caption{\label{graphs}
         Feynman graphs contributing to 
         $e^+e^-\to$ four jets at LO}}
\end{center}
\end{figure}

Invoking the example of four jet--production we will now explain in some 
detail the single steps of the matching procedure used by \APA. The relevant
graphs are depicted in Fig.\,\ref{graphs}.

{\em 1. Step : Choice of number of jets and flavours}

Presuming that we have chosen a sensible $y_{\rm crit}$ for the
evaluation of the matrix elements we are able to choose the number of jets
according to the rates given above, Eq.\,\ref{Rates}. Assume that we are left
with four jet--production, than two possible final states are

\bea
e^+e^-\to q\bar qq'\bar q'\;\;\mbox{\rm and}\;\;\; 
e^+e^-\to q\bar qgg\;,
\eea

which do not mix and can therefore be chosen according to their relative
cross sections. Therefore we will consider in the following the latter 
combination $q\bar qgg$ only.
 
{\em 2. Step : Choice of a specific parton history}

Within the framework of Monte--Carlo methods aiming merely at the correct
average it is perfectly justified to choose now one of the five remaining 
topologies to provide the partons with virtual masses and to account for the
correct colour statistics. Various possibilities exist for this choice, 
``a winner takes it all''--strategy with regard to the relative probabilities 
of the individual topologies encountered as well as an equal probability
distribution between the five diagrams. Within \APA, however, we choose the 
diagram according to the relative probabilities. 

In principle there are various possibilities to define relative probabilities
of Feynman--diagram like topologies. Within \APA we have implemented two.
First, if available, the probabilities ${\cal P}_i$ of each of the five 
topologies can be defined as the squares of the corresponding subamplitudes 
${\cal M}_i$, namely

\bea\label{recon2}
{\cal P}_i = \left|{\cal M}_i\right|^2\,.
\eea

\begin{figure}
\begin{center}
\mbox{\epsfxsize=6cm\epsfysize=6cm\epsffile{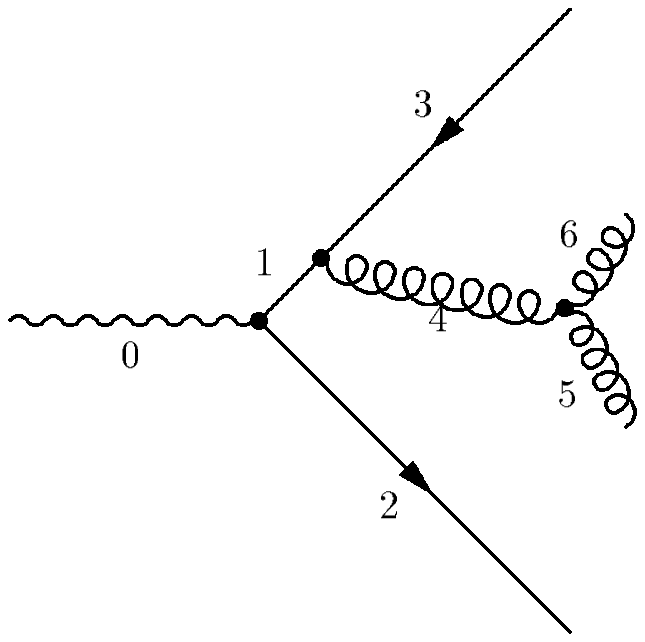}}
\parbox[t]{10cm}{\caption{\label{exgraph}
         Typical graph for $e^+e^-\to$ four jets at LO}}
\end{center}
\end{figure}

The second possibility applies for example for {\tt DEBRECEN} where the 
individual subamplitudes are not supplied. Then, the relative probabilities 
are reconstructed using the parton shower in the fashion of 
\cite{MATCHING}. Consider the topologies depicted in 
Fig.\,\ref{exgraph}. Its relative probability ${\cal P}$ can be defined as

\bea\label{recon}
{\cal P}={\cal P}_{1\to 34}{\cal P}_{4\to 56}=
\frac{1}{t_1}P_{qg}(z_{34})\,\,\frac{1}{t_4}P_{gg}(z_{56})\,,
\eea

with the $P(z)$ the well--known splitting functions, the $t_i$ the squares
of the corresponding four momenta and the $z_i$ the usual energy fractions.

{\em 3. Step : Providing virtual masses}
Having chosen a specific topology it is easy to supply the outgoing partons
with virtual mass invoking the parton shower picture as determined by the 
Sudakov form factor. The starting virtuality for each evolution downwards
is given by the kinematics of the topology. for example in 
Fig.\,\ref{exgraph}, the virtual mass of parton 4 is given by summing and
squaring the known four--momenta of partons 5 and 6 and this virtuality
$t_4$ is employed to determine $t_5$ and $t_6$. Both of them are subject to 
the condition, that no further jets are produced by any subsequent branching 
of them. The same procedure yields virtual masses to any $q\bar q$ pair
in two jet production, where the starting scale is given by the
invariant mass of the intermediate vector boson.

{\em 4. Step : Correcting the kinematics}
Since we want to guarantee four--momentum conservation, the only task left 
is to account for the slight changes in the kinematics due to the fact, that 
the outgoing partons now have acquired a virtual mass. Considering 
subkinematics $a\to bc$ the corrected four momenta $p_i^{\rm cor.}$ are 
given by  

\bea\label{Kincor}
p_{b,c}^{\rm cor.} = p_{b,c}^{(0)} \pm\left(r_cp_c^{(0)}-r_bp_b^{(0)}\right)\,,
\eea

where for the various $r_i$ one has to encounter the following two cases:

\begin{enumerate}
\item{Case 1: b is an internal line, c is outgoing.
      \bea\label{Kincor10}
      r_b &=& \frac{t_a+(t_c-t_b)-\lambda}{2t_a}\,,\nnb\\
      r_c &=& \frac{t_b(t_b-t_c+\lambda)-t_a(t_a-t_c-\lambda)}
               {2t_a(t_b-t_a)}\,.
      \eea}
\item{Case 2: b and c are outgoing.
      \bea\label{Kincor11}
      r_{b,c} &=& \frac{t_a \pm(t_c-t_b)-\lambda}{2t_a}\,. 
      \eea}
\end{enumerate}

Obviously, not only the virtual masses are provided. Additional changes 
alter slightly the $z$ and tend to narrow the angels $\theta_{bc}$.
However, a careful study, like for instance of the various four--jet 
correlation angles \cite{4JETth} showed that these are only minor changes.

\section{Results}

We have performed a comparison of a variety of observables at a c.m. 
energy of $91$ GeV at the level of matrix elements, parton showers
and hadrons using {\tt PYTHIA} \cite{PYTHIA}, {\tt HERWIG} \cite{HERWIG} 
and our event generator {\tt APACIC++}. for the latter we used matrix 
elements for the production of up to five jets via QCD provided by \AME.
For \APA we employed the string hadronization \cite{STRING} in the form 
of \cite{LUND} by linking the corresponding routines of JETSET to our code. 
We did not take into account any initial state radiation.

We would like to divide the presentation and discussion of results into
two parts, one part flashing over some representative event--shape observables
and the like, proving clearly, that \APA is perfectly capable to reproduce
the experiment. In the other part we will restrict ourselves to the parton 
level only and show, that our matching formalism has some clear benefits
describing the topological structure of four jet--events.

\subsection{Comparison with event--shapes}

\begin{figure}
\mbox{\begin{tabular}{cc}
\epsfxsize=6cm\epsfysize=4cm\epsffile{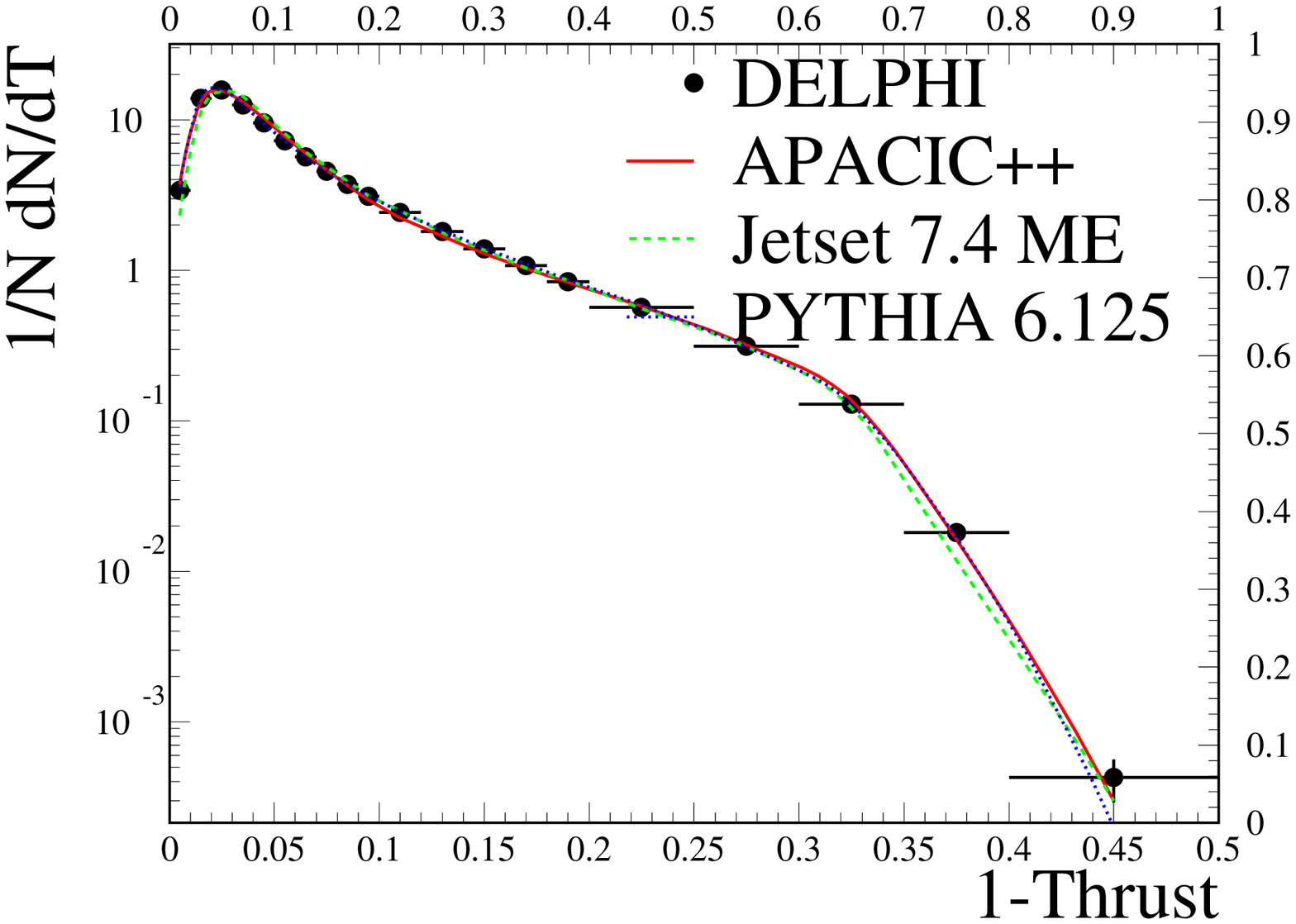}&
\epsfxsize=6cm\epsfysize=4cm\epsffile{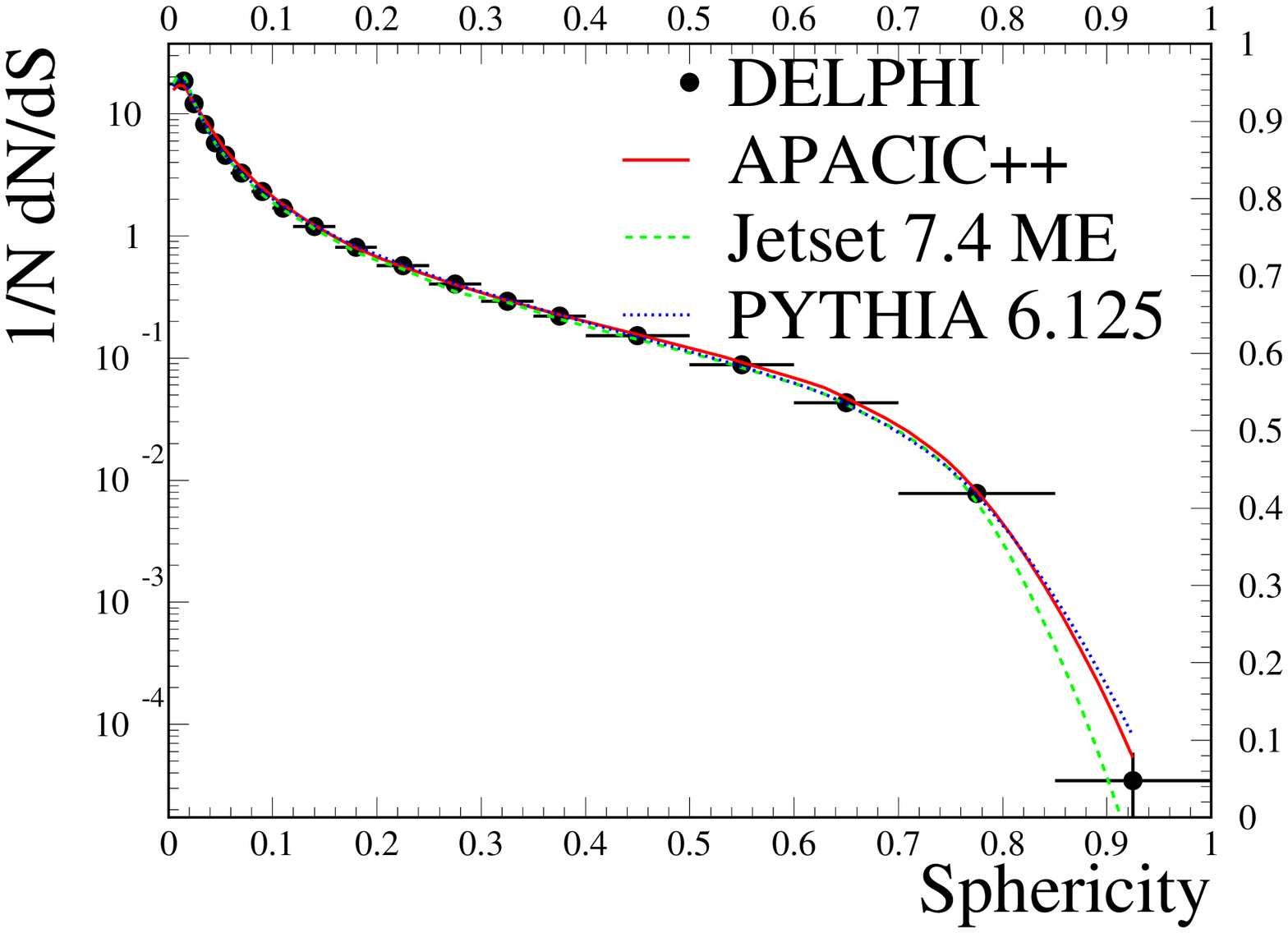}\\
\epsfxsize=6cm\epsfysize=4cm\epsffile{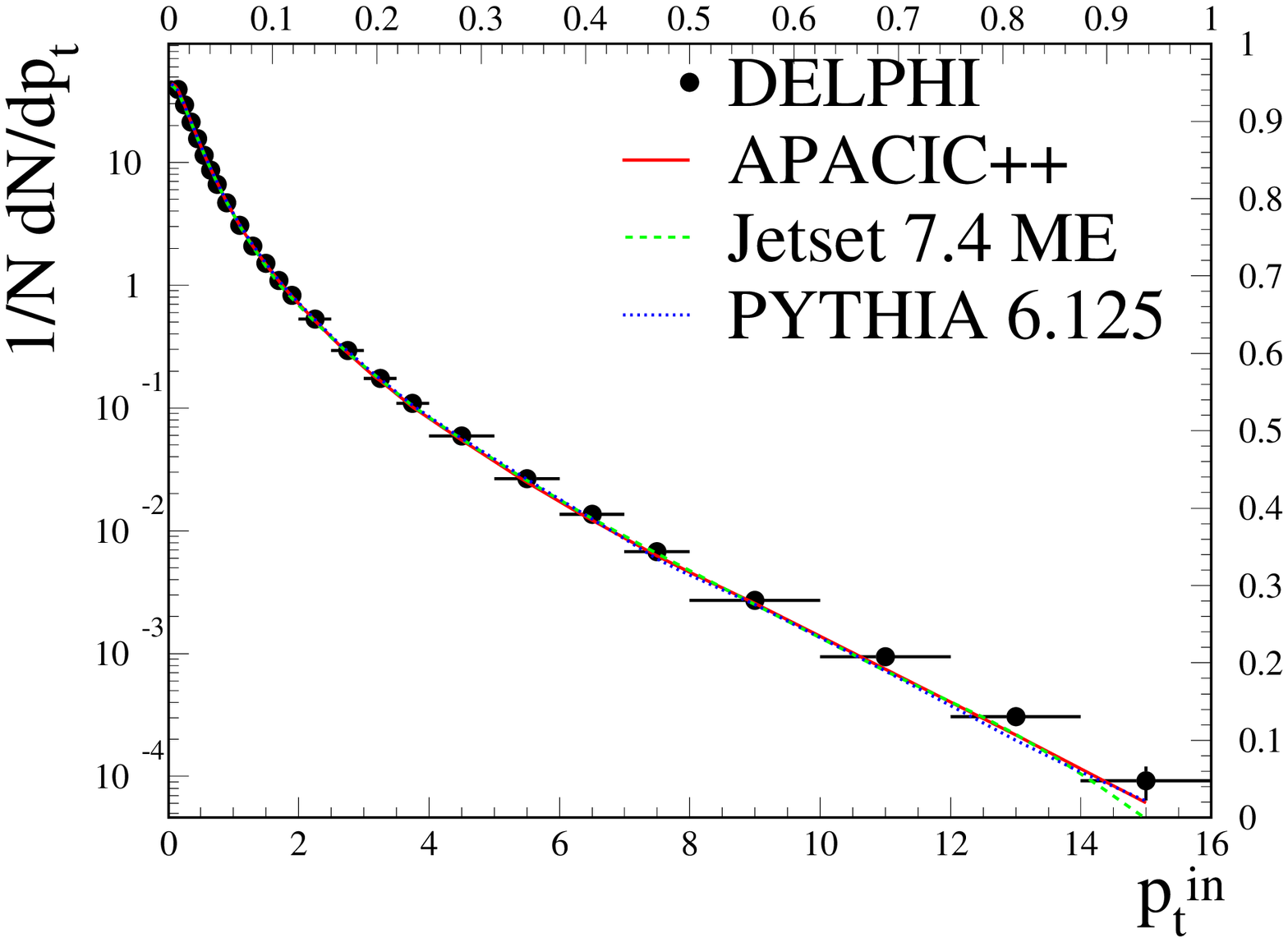}&
\epsfxsize=6cm\epsfysize=4cm\epsffile{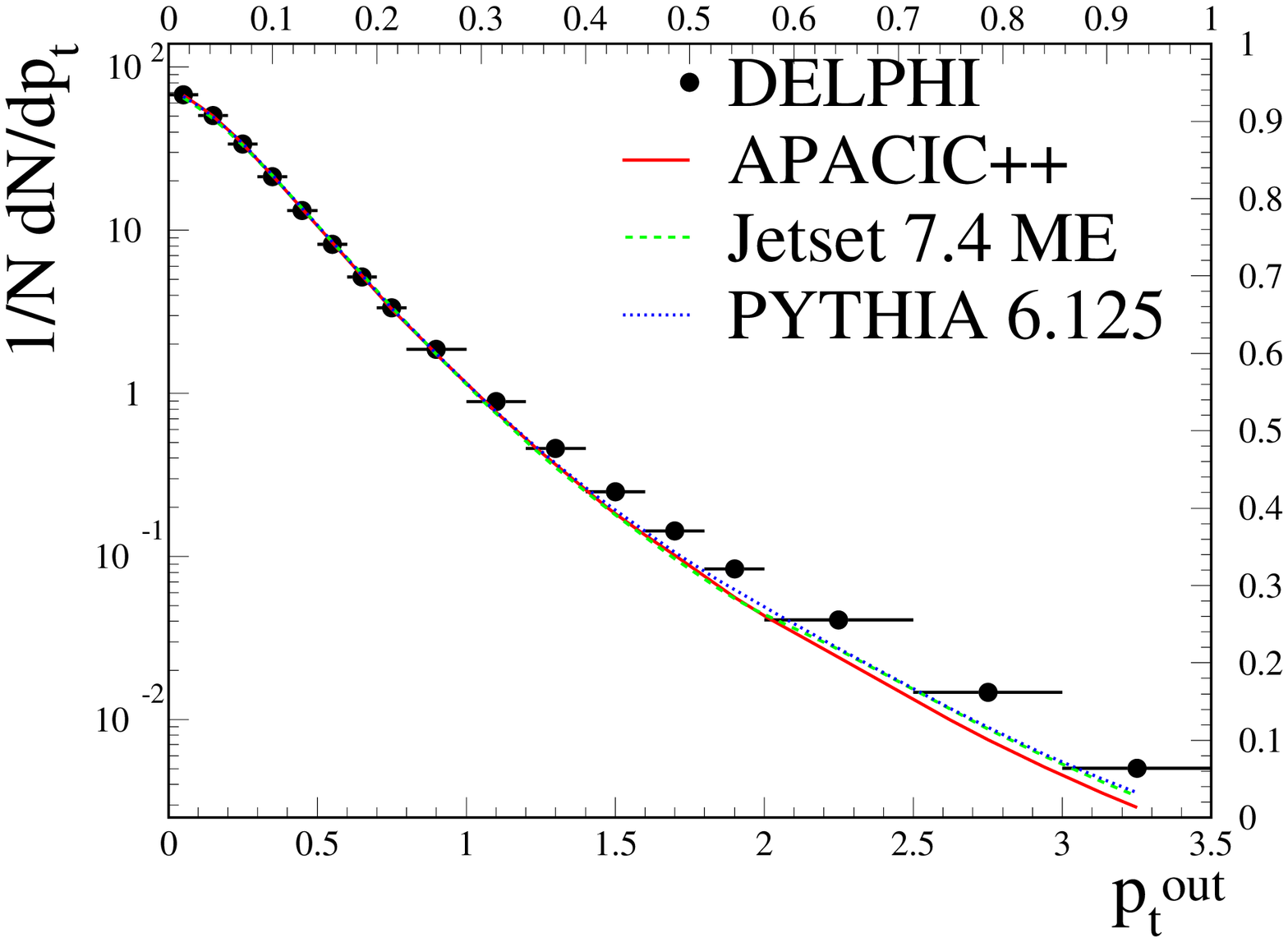}\\
\epsfxsize=6cm\epsfysize=4cm\epsffile{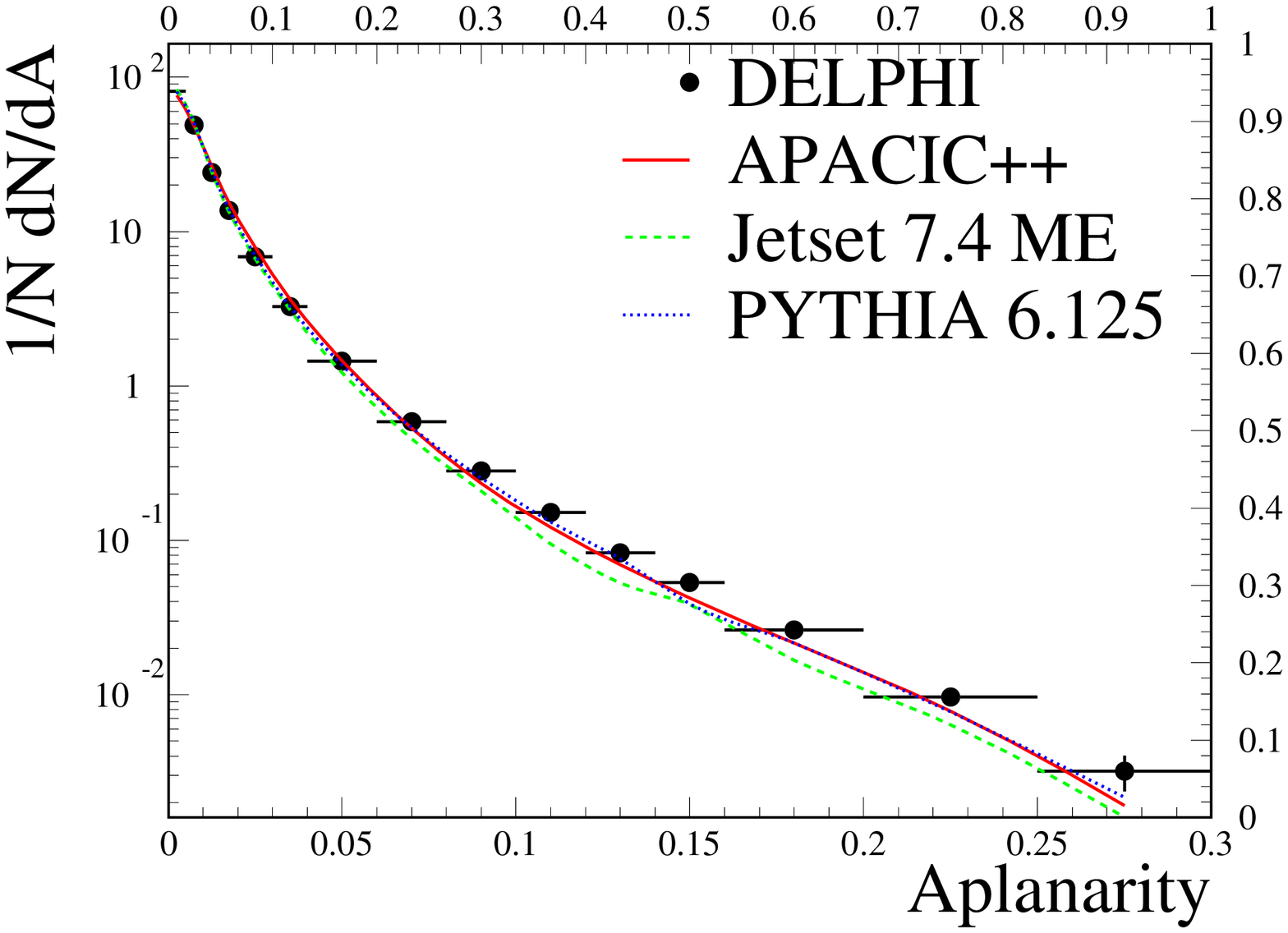}&
\epsfxsize=6cm\epsfysize=4cm\epsffile{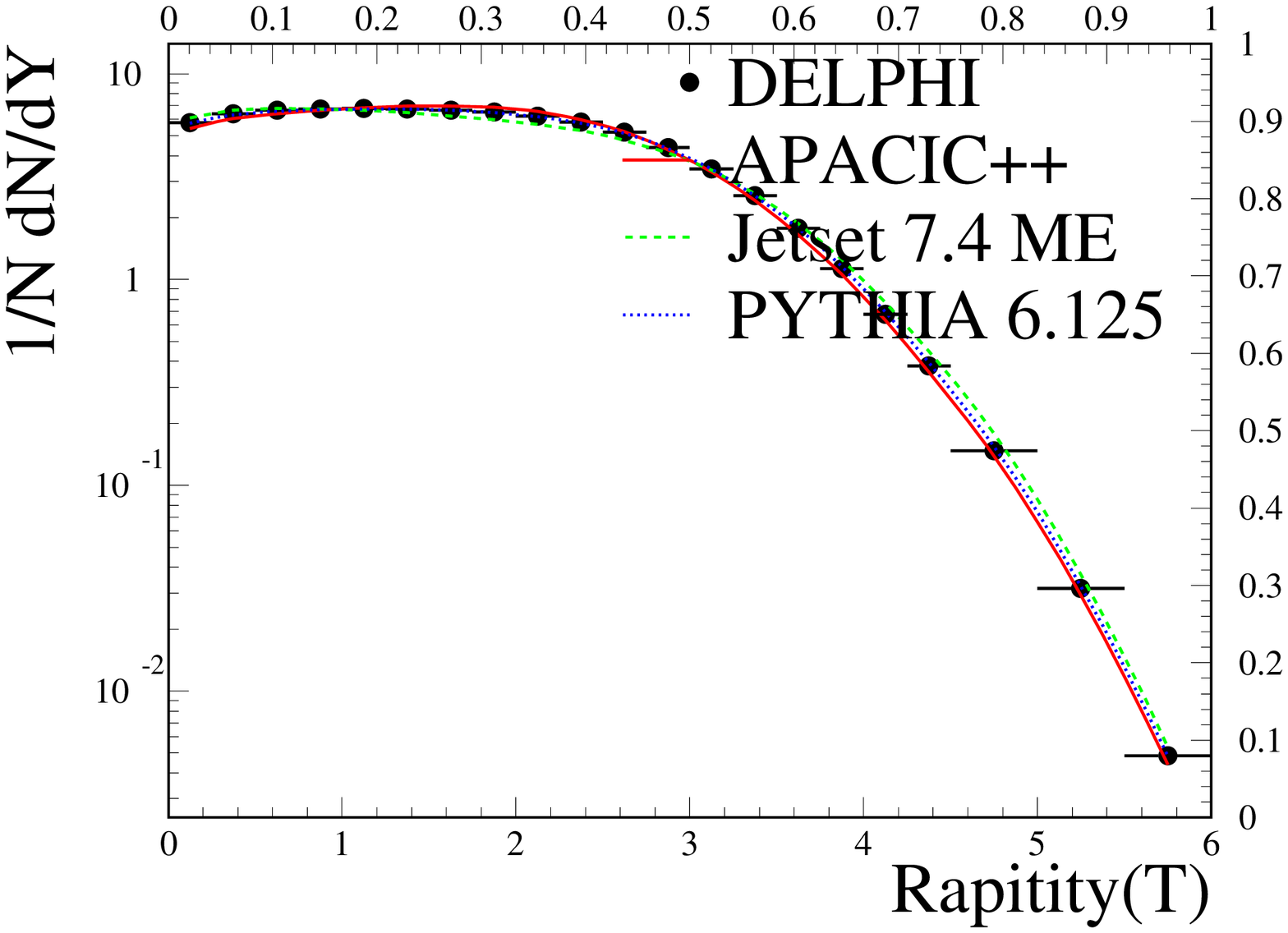}\\
\end{tabular}}
\parbox[t]{13.5cm}{
\caption{\label{observs1} Comparison of experimental data and 
event generators for a variety of event shape observables at the hadron level
at the $Z$--pole. We employed the Lund--String hadronization of {\tt PYTHIA}
for {\tt APACIC++}.} The plots stem from \cite{Wupps1} utilizing data
from \cite{Wupps2}.}
\end{figure}

Comparing results of \PYT and \APA with each other and experimental data
provided by the DELPHI--collaboration we found an encouraging agreement 
for most of the observables. For \PYT we employed as an additional channel we 
denote by {\tt JETSET} the matrix elements provided there with subsequent 
hadronization without intermediate parton shower.
For a representative extract of various event
shape observables see Fig.\,\,\ref{observs1}. There, we depict the 
sphericity distribution, the $1-$thrust distribution as well as the inner 
and outer transversal momentum distribution. Additionally we depict
the aplanarity and the rapidity with respect to the thrust axis.
Obviously the results of \APA are in pretty good agreement with data 
indicating that our approach to match matrix elements and parton showers 
describes reasonably the interplay of various numbers of jets as well as the 
overall features of $e^+e^-$ events.

We would like to introduce briefly the observables we display to the
reader unfamiliar with them. For this purpose we consider first the
tensor constructed out of the three--momenta $p$ of final--state particles,

\bea
S^{\alpha\beta} = \frac{\sum_i (p_i^{\alpha}p_i^{\beta})}{\sum_i p_i^2},
\quad \alpha,\beta = 1,2,3\,,
\eea

with eigenvalues $\lambda_{1,2,3}$. The combination

\bea
S = \frac{3}{2}(\lambda_1 + \lambda_2)
\eea

of the two smalles eigenvalues defines the sphericity. In contrast, 
aplanarity is given by

\bea
A=\frac32\,\lambda_1\,.
\eea

Thrust is defined by the maximal value of

\bea
T=\mbox{\rm max}_{\vec n}\left(
\frac{\sum\limits_i|\vec p_i\vec n|}{\sum\limits_i|\vec p_i|}\right)\;,
\eea

where $\vec n$ is a free vector to be chosen accordingly. The vector
$\vec n$ yielding the maximal value of $T$ is the thrust--axis. 

$p_\perp^{\rm in}$ and $p_\perp^{\rm out}$ are the components of the 
transversal momenta being inside the event--plane or perpendicular.
The rapidity here is taken with respect to the thrust--axis.

\subsection{Topological structure of four jet--events}

However, the validity of our matching procedure can be verified in more 
depth considering the topological structure of multijet--events as exemplified
by four--jet events. Ordering the jets by their energies, 
$E_1\ge E_2\ge E_3\ge E_4$, typical observables describing these processes 
are the modified Nachtmann--Reiter--, the Bengtson--Zerwas-- and the
K{\"o}rner--Schierholz--Willrodt--angle as well as the angle
$\alpha_{34}$ between the two least energetic jets \cite{4JETth,ESW96},

\bea\label{angledef}
\theta_{\rm NR}^* &=& \angle(\vec{p_1}-\vec{p_2},\vec{p_3}-\vec{p_4})
                      \,,\nnb\\
\chi_{\rm BZ} &=& \angle(\vec{p_1}\times\vec{p_2},\vec{p_3}\times\vec{p_4})
                      \,,\nnb\\
\Phi_{\rm KSW} &=& \angle(\vec{p_1}\times\vec{p_3},\vec{p_2}\times\vec{p_4})
\,.
\eea

\begin{figure}
\begin{center}
\mbox{\begin{tabular}{cc}
\epsfxsize=6cm\epsffile{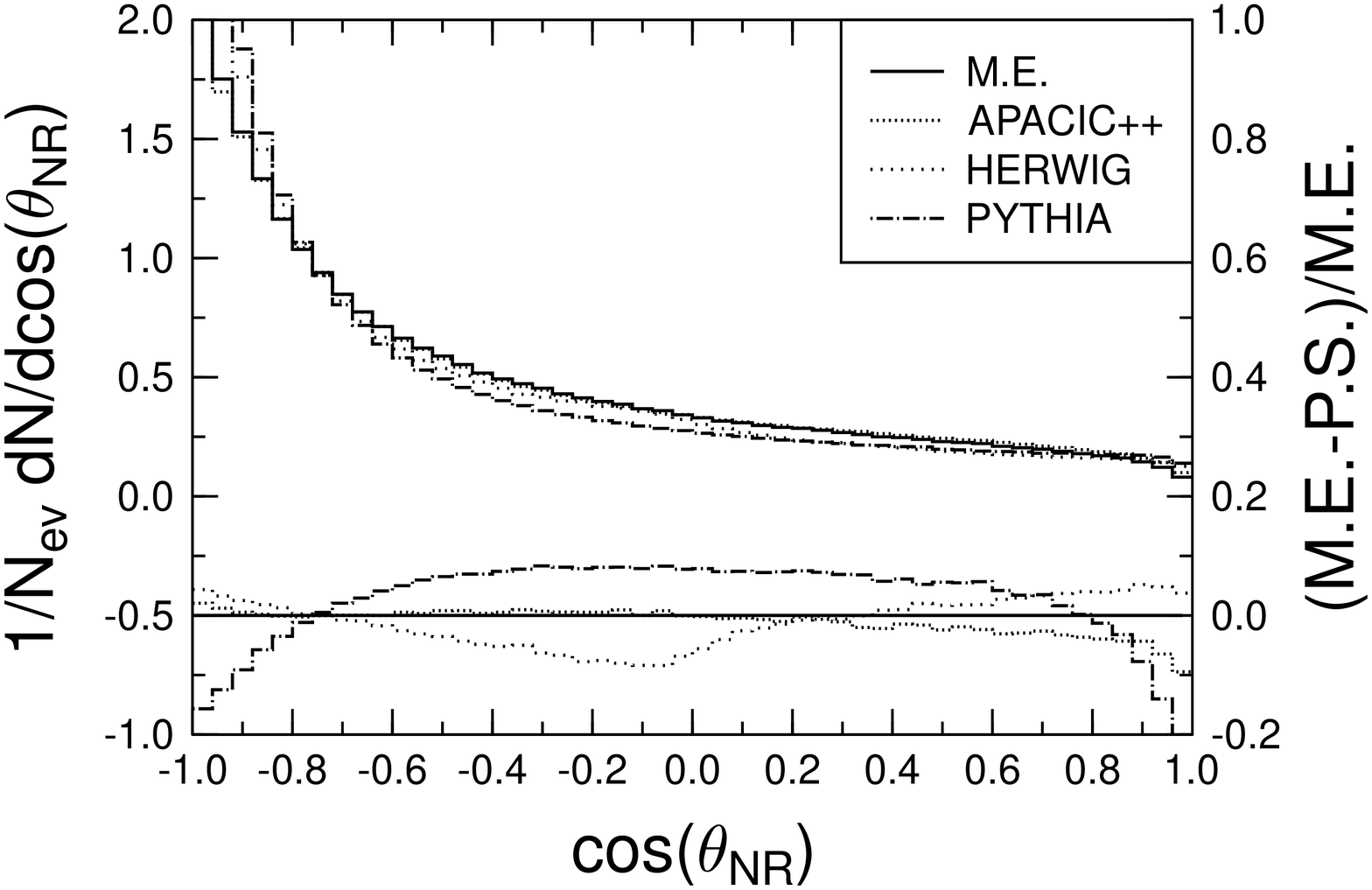}&
\epsfxsize=6cm\epsffile{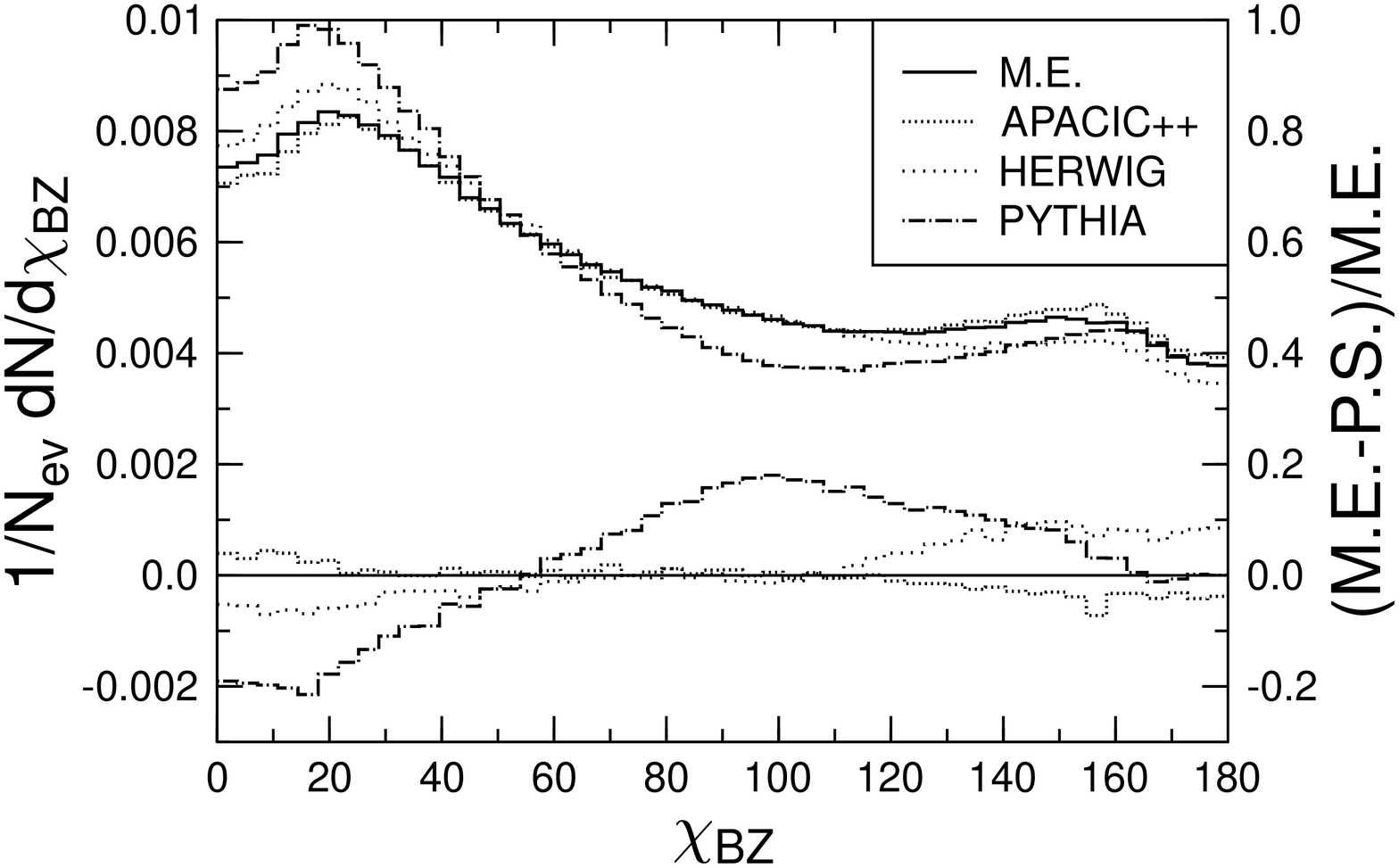}\\
\epsfxsize=6cm\epsffile{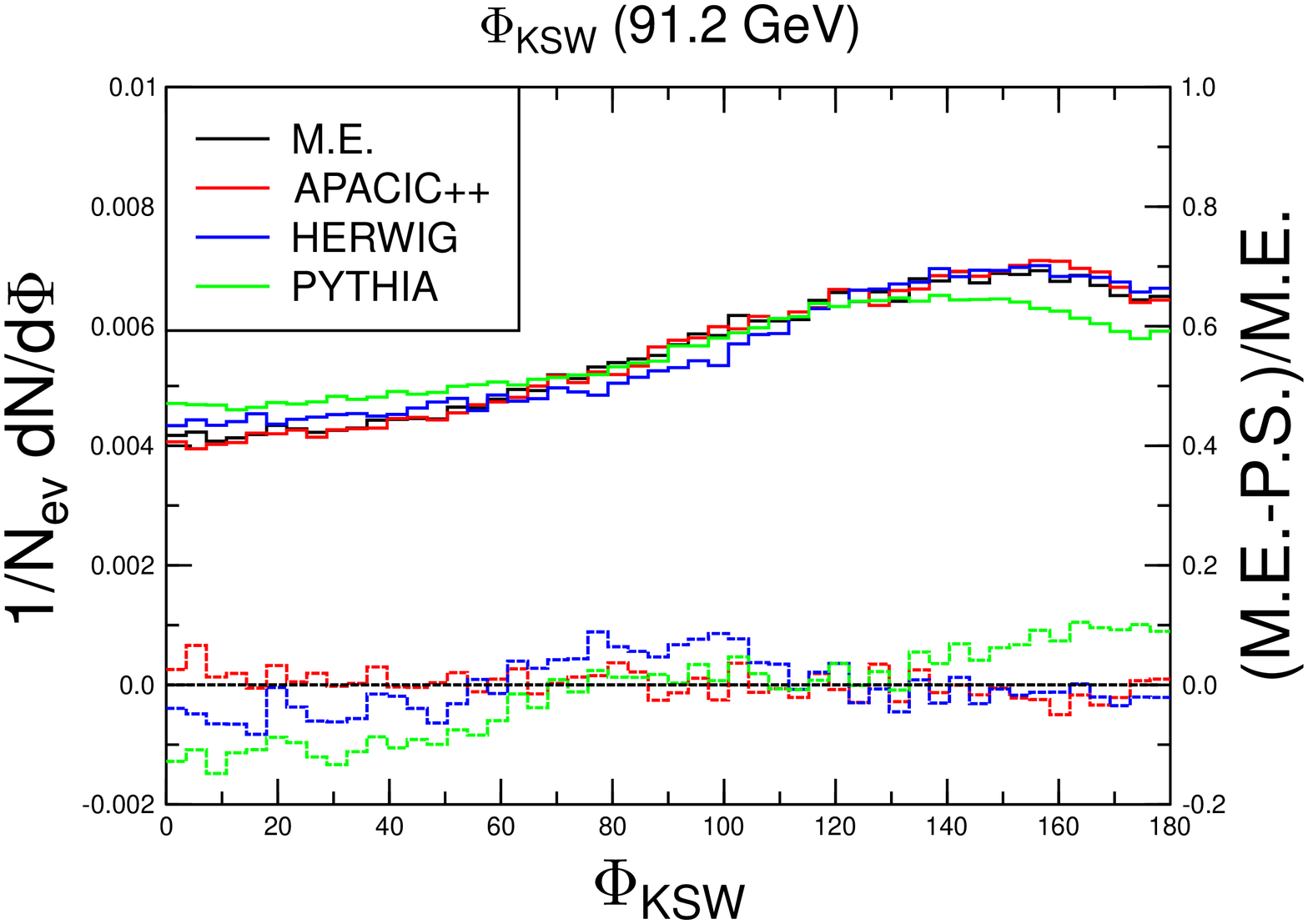}&
\epsfxsize=6cm\epsffile{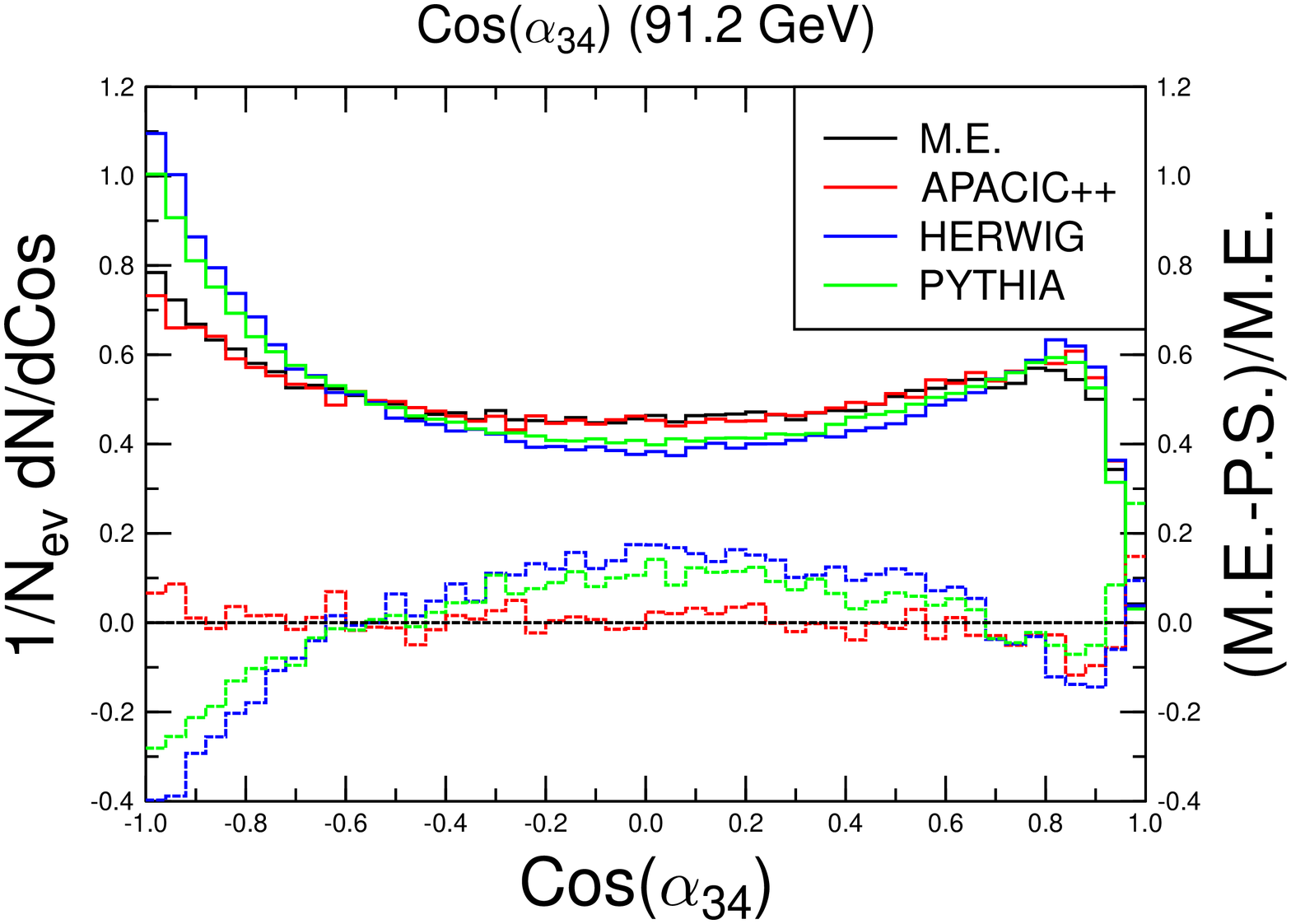}
\end{tabular}}
\parbox[t]{13.5cm}{\caption{\label{angles}
         Distributions for the modified Nachtmann--Reiter--, the
         Bengtson--Zerwas--, the K{\"o}rner--Schierholz--Willrodt--angle 
         and for $\alpha_{34}$ as given in Eq. \ref{angledef} obtained
         by the various event generators. 
         For the definition of jets the Durham--scheme with 
         $y_{\rm cut}=0.002$ was employed for all final states
         as well as for the matching of the matrix elements and
         the parton shower. The upper lines show the corresponding
         differential rates with respect to the numbers on the left 
         axis whereas the errors relative to the matrix element
         expression, namely (M.E.-P.S.)/M.E 
         are given by the appropriate lower lines with respect
         to the numbers on the right axis.}}
\end{center}
\end{figure}

In Fig.\,\,\ref{angles} we display the angular distributions of the partons 
after the shower generated by the various event generators in comparison to the
distributions resulting from the corresponding matrix elements. 

\section{Conclusions}

Obviously APACIC++ is perfectly capable to describe in a precise and reliable manner
the four jet topologies. Therefore one is tempted to conlude, that the parton shower 
and the matrix elements are matched appropriately. The few sizeable deviations 
of the topologies at the parton shower level from the matrix elements
are collimated in the region of nearly collinear jets. This is not too surprising, 
however, since the jet evolution softens the initial partons to jets and widens
them to jet--cones which in turn may easily overlap. Of course this alters the results
slightly. However, in principle, this exactly reflects the picture employed of 
hard produced partons widening to jets. We therefore conclude, that the matching 
succeeded.

In contrast, the two other event generators considered at the present state
do not include an accurate matching procedure for four--jet events.
Therefore their failure in describing such topologies consistently at the parton level 
merely reflects the fact, that the angular structures of four--jet events are due
to correlations not embedded in the parton shower like for instance interferences 
of single diagrams. 

On the other hand, it should be noted, that all of the event generators displayed here
reproduce the overall features of $e^+e^-$--annihialtions into hadrons in a fairly 
satisfying manner, even though the intrinsic parameters of our code \APA have not 
been fully tuned.

Summarizing we would like to state, that we have proposed a obviously working
general approach to match parton showers and arbitrary matrix elements in the 
framework of QCD event generators. We have implemented this ansatz into the newly 
developped event generator \APA, which linked with \AME will offer new possibilities 
to describe precision data concerning multijet--events at LEP II and beyond.

\section*{Acknowledgements}

We are deeply indebted to J. Drees, K. Hamacher and U. Flagmeyer for helpful 
discussions concerning various aspects of $e^+e^-$--collisions at LEP and
for some of the result plots displayed. Additionally, we would like to 
thank B. Iv{\'a}nyi for a large number of lively and clarifying discussions. 
We gratefully acknowledge financial support by Minerva, DFG, BMBF and GSI. 

Both codes, \APA and \AME can be obtained upon request from 
F. Krauss \thanks{E-mail: krauss@physics.technion.ac.il} and 
R. Kuhn  \thanks{E-mail: kuhn@theory.phy.tu-dresden.de}.

\end{document}